\documentclass[preprint,amsmath,amssymb,nofootinbib]{revtex4}
\usepackage[utf8]{inputenc}
\usepackage{rotating}
\usepackage{hyperref}
\usepackage{array}
\usepackage{subfig}
\usepackage{float}
\usepackage{fancyhdr}
\usepackage{slashed}
\usepackage[inline]{enumitem}
\usepackage{dcolumn}
\usepackage{booktabs}
\usepackage{makecell}
\usepackage{placeins}
\usepackage[export]{adjustbox}
\usepackage{bm}
\usepackage{xcolor}
\usepackage{graphicx}
\usepackage{diagbox}
\usepackage{multirow}
\usepackage{ctable}
\usepackage[normalem]{ulem}
\usepackage[T1]{fontenc}
\definecolor{orange}{rgb}{1,0.5,0}

\begin{document}

\title{Future collider sensitivities to $\nu$SMEFT interactions}

\author{Luc\'{\i}a Duarte}
\email{lucia.duarte@fcien.edu.uy}
\affiliation{Instituto de F\'{\i}sica, Facultad de Ciencias,
 Universidad de la Rep\'ublica \\ Igu\'a 4225, (11400) 
Montevideo, Uruguay.}

\author{Daniel Chalençon Maisian}
\affiliation{Universitat de València, Burjassot, (46100) València, Spain.}

\author{Tom\'as Urruzola}
\affiliation{Instituto de F\'{\i}sica, Facultad Ingenier\'ia,
 Universidad de la Rep\'ublica \\ Julio Herrera y Reissig 565, (11300) 
Montevideo, Uruguay,}
\affiliation{Instituto de F\'{\i}sica, Facultad de Ciencias,
 Universidad de la Rep\'ublica \\ Igu\'a 4225,(11400) 
Montevideo, Uruguay.}

\begin{abstract}

The discovery of neutrino oscillations and masses provides strong motivation to extend the Standard Model by including right-handed neutrinos, which lead to heavy neutrino states that could exist at the electroweak scale. These states may also be influenced by new high-scale, weakly interacting physics. Incorporating right-handed neutrinos into an effective field theory framework -the $\nu$SMEFT- offers a systematic approach to study the phenomenology of heavy neutrinos in current and upcoming experiments.
In this work, we present the first prospective 95\% exclusion plots achievable at a future lepton collider operating at a center-of-mass energy of $\sqrt{s}=0.5 ~\rm{TeV}$ for what we term the agnostic $\nu$SMEFT scenario. This study focuses on the high-mass regime where the heavy neutrino $N$ decays promptly into leptons and jets. Specifically, we analyse the processes $e^+e^- \to \nu N \to \nu \mu^{-}  \mu^{+} \nu$ and $e^+e^- \to \nu N \to \nu \mu^{-} \mathrm{j} \mathrm{j}$, deriving the exclusion regions in the $\frac{\alpha}{\Lambda^2}$ vs. $m_N$ parameter space. When compared to prospective limits for the LHeC, we find that the semi-leptonic process with final jets in a lepton collider offers the greatest sensitivity, even with a straightforward cut-based analysis. The expected bounds are as stringent as those considered in recent studies for the low-mass regime where the $N$ may be long-lived and detectable via displaced decay searches, both at the LHC and future colliders.

\end{abstract}

\maketitle

\section{Introduction}{\label{intro}}
 

The presence of light neutrino masses and the phenomenon of neutrino oscillations can be explained within a minimal extension of the Standard Model (SM) Lagrangian by introducing sterile right-handed neutrinos $N_R$, which enable a lepton-number-violating Majorana mass term, as in the Type-I seesaw mechanism \cite{Minkowski:1977sc, Mohapatra:1979ia, Yanagida:1980xy, GellMann:1980vs, Schechter:1980gr}. The Majorana mass scale, a parameter independent of the electroweak symmetry breaking, is typically assumed to be large in the naive (high-scale) seesaw framework. This results in the suppression of the induced mass of the -predominantly active- light neutrinos, while generating very heavy massive states alongside the light ones. Alternatively, the small masses of the known neutrinos could arise from symmetry-based arguments, as proposed in the linear and inverse seesaw models \cite{Malinsky:2005bi, Mohapatra:1986bd}, which lower the mass scale of the heavy neutrinos and make their phenomenology testable at laboratory energies. Even if these heavy neutral leptons (HNLs) or heavy neutrinos $N$ were experimentally accessible, their interactions with SM particles are suppressed due to their small mixing $U_{\ell N}$ with active neutrinos $\nu_{\ell L}$ which are strongly constrained by experimental limits \cite{Abdullahi:2022jlv}. As a result, these interactions would be exceedingly weak and likely undetectable.

Nonetheless, various forms of new physics could exist at energies significantly above the electroweak scale. Their effects on SM degrees of freedom are systematically analyzed using the Standard Model Effective Field Theory (SMEFT) framework. If heavy neutrinos are sufficiently light to be included in the low-energy spectrum, such new physics could influence their behaviour and likely dominate over interactions arising from their mixing with active neutrinos. Consequently, the interactions of HNLs with SM particles may reflect the remnants of ultraviolet (UV) physics, which can be described by an effective field theory where the HNLs themselves are included as fundamental components. This framework extends the SMEFT to incorporate right-handed neutrinos and is referred to as $\nu$SMEFT\footnote{It is also known in the literature as SMNEFT, $N_R$SMEFT and $\nu_R$SMEFT.}\cite{Anisimov:2006hv,Graesser:2007yj,Graesser:2007pc,delAguila:2008ir,Aparici:2009fh,Liao:2016qyd,Bhattacharya:2015vja,Li:2021tsq}.

The $\nu$SMEFT can parametrize the effect of a plethora of existing -or yet to be imagined- UV physics models in experiments performed at electroweak scale energies, where the heavy $N$ can be treated as an accessible degree of freedom with a mass $m_N$ in the hundred-GeV range. In this mass window, the heavy neutrino with effective interactions, while being accessible for direct production, can also decay promptly into on-shell top quarks, Higgs bosons, and the electroweak standard vector bosons too, leading to final states commonly studied for standard and BSM interactions at the LHC and future lepton and electron-proton colliders. 

Some well-known UV-complete models that generate $\nu$SMEFT operators at the electroweak scale include the Left-Right Symmetric Model \cite{Mohapatra:1979ia, Mohapatra:1980yp}, models where right-handed neutrinos interact with $Z'$ bosons \cite{Mohapatra:1980qe} or with leptoquarks \cite{Bhaskar:2023xkm, Cottin:2021tfo}, and scenarios that combine several of these interactions \cite{Das:2016vkr}.
 
Initially proposed by the authors of \cite{delAguila:2008ir} as a dedicated EFT for studying neutrino interactions, the $\nu$SMEFT framework has since seen significant advancements in both theoretical and phenomenological aspects. Our research group has contributed to study the possible phenomenology of the $\nu$SMEFT operators, initially with the calculation of the full set of $N$ decay channels \cite{Duarte:2015iba, Duarte:2016miz}, with signatures at different colliders in \cite{Duarte:2014zea, Duarte:2016caz, Duarte:2018xst, Duarte:2018kiv, Zapata:2022qwo, Zapata:2023wsz}, at neutrino telescopes \cite{Duarte:2016smd}, and also in the context of $B$ meson decays \cite{Duarte:2019rzs, Duarte:2020vgj}. While the $\nu$SMEFT framework started to gain attention, the phenomenology given by particular operators involving Higgs boson interactions was studied in \cite{Caputo:2017pit, Butterworth:2019iff, Jones-Perez:2019plk, Barducci:2020icf}, and multiple prospects for new collider signatures from heavy neutrinos with effective interactions, both with prompt and displaced decays were studied in \cite{Cottin:2021lzz, Beltran:2021hpq, Barducci:2022hll, Mitra:2022nri, Mitra:2024ebr, Biswas:2024gtr, Beltran:2025ilg}. Heavy neutrinos with lower mass can be produced in meson or tau lepton decays: this feature has been explored in the $\nu$SMEFT context in \cite{Yue:2018hci, DeVries:2020jbs, Zhou:2021ylt, Beltran:2022ast, Beltran:2023nli}. General neutrino interactions were also studied from the EFT perspective in \cite{Bischer:2019ttk}, focusing on constraints from low-energy observables and pointing to possible UV completions involving leptoquarks. The relation to neutrinoless double beta decay was studied in \cite{Dekens:2020ttz,Dekens:2021qch}. The heavy neutrinos dipole momenta, in their interactions with photons, were studied in this context in \cite{Biekotter:2020tbd, Cirigliano:2021peb, Barducci:2022gdv, Beltran:2024twr, Barducci:2024kig} and also bounds were obtained from displaced photons searches at LHC experiments \cite{Delgado:2022fea, Duarte:2023tdw}. Recently, Ref.~\cite{Fuyuto:2024oii} studied the generation of $\mathcal{O}(\text{keV})$ sterile neutrino dark matter, while another dark matter scenario within the $\nu$SMEFT framework was explored in \cite{Borah:2024twm}.

The one loop renormalization group evolution of $\nu$SMEFT operators was tackled in \cite{Chala:2020vqp, Chala:2020pbn, Ardu:2024tzb}, and UV completions were systematically classified in \cite{Beltran:2023ymm}, while explicit matching with the minimal Left-Right symmetric model was done in \cite{deVries:2024mla}. The  available information is starting to be organized in constraints to specific operators \cite{Alcaide:2019pnf, Fernandez-Martinez:2023phj} obtained from the reinterpretation of existing experimental searches.  Also, very recent prospects for constraining certain $\nu$SMEFT operators with monophoton searches and displaced vertex decay signatures at the FCC-ee can be found in \cite{Bolton:2025tqw}. 

Most of the above mentioned collider $\nu$SMEFT phenomenology literature has focused on analyses involving specific subsets of the effective operators, chosen to contribute to the signals investigated in each particular study, and sometimes restricting to contributions of only certain quark and lepton flavors to enhance the sensitivity to the chosen signals and avoid existing bounds. For example, different operators are chosen to produce the heavy $N$ and to induce its decay. This treatment oversimplifies the possible underlying BSM scenarios and might not be realistic. One option is to restore the studies to specific UV-models that can be precisely matched to the $\nu$SMEFT operators, in a top-down perspective \cite{deVries:2024mla}. Also, prospective studies can be performed by considering the sensitivity to a single operator acting at a time.\cite{Bolton:2025tqw}. While being a standard procedure, this approach does not take into account that the UV physics leading to one operator can also contribute to other effective interactions when renormalization group running gives place to operators mixing, or even that different UV models could be the joint physics responsible for the observed effects. 

Here we take a complementary approach, considering a simplified benchmark scenario, which we call 'agnostic', including every dimension 6 $\nu$SMEFT operator, and taking all them to have the same numerical value. This agnostic benchmark is intended to provide a simple scenario where prospects can be interpreted, and taken into account by the experimental community \cite{Ahmadova:2025vzd,FCC:2018evy}, in spite of being a worst-case scenario from the point of view of the enhancement of specific signals, as it allows for amplitude cancellations between different $\nu$SMEFT operators and with the SM backgrounds.

In this context, we study the sensitivity projections for the heavy Majorana $N$ in future $e^+e^-$ colliders, focusing on processes where it is singly produced together with a light neutrino and decays into di-muon and semi-leptonic final states: $e^+e^- \to \nu N \to \nu \mu^- \mu^+ \nu$ and $e^{+}e^{-} \to \nu N \to \nu \mu^{-} \mathrm{j} \mathrm{j}$, with $\nu=\nu_{e,\mu,\tau}$ and light flavored jets $\mathrm{j}= u,d,c,s$. The simplified agnostic benchmark scenario allows for the study of a two-dimensional parameter space, providing 95\% CL exclusion limits in the mass-coupling $(m_N, \frac{\alpha}{\Lambda^2})$ plane. We introduce the agnostic $\nu$SMEFT benchmark scenario in Section \ref{sec:agnosticB}, discuss the collider study and the comparison with previous LHeC sensitivities \cite{Zapata:2023wsz} in Section \ref{sec:ColliderSensitivities}, and present our conclusions in Section \ref{sec:summary}.

\section{The agnostic $\nu$SMEFT scenario}\label{sec:agnosticB}

Our starting point is to consider the SM Lagrangian to be extended with only \emph{one} right-handed neutrino $N_R$ with a Majorana mass term ($\sim M_N$). While at least two right-handed $N_R$ states are required to reproduce the measured masses and mixings with light neutrinos, this simplifying assumption retains the main phenomenology and corresponds to scenarios where the additional massive $N$ are too heavy to impact in low-energy observables. The renormalizable $d=4$ Lagrangian extension then reads 
\begin{eqnarray}\label{eq:LagSeesaw}
\mathcal{L}_{\rm d=4}= \overline{N_R} i\slashed{\partial} N_R - \left( \frac{M_N}{2} \overline{N^{c}_R} N_R + \sum_{\ell} Y_\ell ~ \overline{L_\ell}\tilde{\phi} N_R + \text{ h.c.}\right).
\end{eqnarray}
Once diagonalized, this Lagrangian leads to a massive state $N$ as an observable degree of freedom, together with the three known light neutrino states $\nu_i$ (with masses $m_{\nu_i}\sim 0.1$ eV), which are all of Majorana nature. The active flavor $\ell=e,\mu,\tau$ neutrino eigenstates $\nu_{\ell L}$ contain some part of the heavy $N$ due to the mixing $U_{\ell N} = Y_\ell ~v/\sqrt{2} M_N$: 
\begin{equation}\label{eq:mixing}
    \nu_{\ell L}=\sum_{i=1}^{3}U_{\ell i}\nu_i+U_{\ell N}N.
\end{equation}
The heavy state $N$ is in turn predominantly composed of the right-handed state $N \simeq N_R$ with negligible mixing with the active $\ell$ flavor states $\nu_{\ell L}$. This mixing is constrained by the naive seesaw relation $U_{\ell N}\lesssim \sqrt{ \frac{m_{\nu}}{M_{N}}}$, resulting in suppressed interactions with the SM electroweak currents for $N$ masses above the GeV scale.

\renewcommand{\arraystretch}{1.2}
\begin{table}[tbp]
\begin{adjustbox}{width=\textwidth,center} 
\centering
\begin{tabular}{| >{\arraybackslash}p{3 cm}| l l |>{\arraybackslash}p{6 cm}| l |}
\firsthline
\textbf{Type}      &     \textbf{Operator}       &            & \textbf{Interactions}   & \textbf{Coupling}   \\ \hline\hline
$N$  mass $d=5$ & $\mathcal{O}^{d=5}_{N\phi}$ ($\mathcal{O}^{d=5}_{\rm Higgs}$)  & $(\bar{N}N^{c})(\phi^{\dagger} \phi)$  &  $ ~h N N$ and Majorana mass term  & $\alpha^{d=5}_{N\phi}$   \\ \hline 
Dipole $d=5$ & $\mathcal{O}^{(5)}_{NB}$  & $(\bar{N}_a \sigma_{\mu \nu} N^{c}_b) B^{\mu \nu}$, $a \neq b$ & $~$ Dipoles $~d_{\gamma}, d_Z$ & $\alpha^{d=5}_{NB}$ \\  \hline \hline
$h$-dressed mixing   & $\mathcal{O}^{(i)}_{LN\phi}$ ($\mathcal{O}_{\rm LNH}^\beta$)  & $(\phi^{\dag}\phi)(\bar L_i N \tilde{\phi})$  & $~$ Yukawa$+$doublet ($U_{\ell N}$. and $m_{\nu}$) & $\alpha^{(i)}_{LN\phi}$  \\ \hline 
$~$ Bosonic & $\mathcal{O}_{NN\phi}$ ($\mathcal{O}_{\rm HN}$) & $ i(\phi^{\dag}\overleftrightarrow{D_{\mu}}\phi)(\bar N \gamma^{\mu} N)$                                & $~$ Neutral current ($NNZ$) & $\alpha_{NN\phi}=\alpha_{Z}$ \\ 
$~$ Currents   & $\mathcal{O}^{(i)}_{Nl\phi}$ ($\mathcal{O}_{\rm HN\ell}^{\beta}$) & $i(\phi^T \epsilon D_{\mu}\phi)(\bar N \gamma^{\mu} l_i) $  & $~$ Charged current ($N l W$) & $\alpha^{(i)}_{Nl\phi}=\alpha^{(i)}_{W}$ \\  \hline
$~$ Dipoles & $\mathcal{O}^{(i)}_{NB}$ ($\mathcal{O}_{\rm NB}$)  & $(\bar L_i \sigma^{\mu\nu} N) \tilde \phi B_{\mu\nu}$  &  $~$ One-loop level generated & $\alpha^{(i)}_{NB}/(16\pi^2)$ \\ 
$ $  & $\mathcal{O}^{(i)}_{NW}$ ($\mathcal{O}_{\rm NW}^\beta$)    & $(\bar L_i \sigma^{\mu\nu} \tau^I N) \tilde \phi W_{\mu\nu}^I$ & $~~~d_{\gamma} , d_Z, d_W$ & $\alpha^{(i)}_{NW}/(16\pi^2)$ \\ \hline 
$ $  & $\mathcal{O}^{(i)}_{QNN}$ ($\mathcal{O}_{\rm QN}$)  &   $(\bar{Q_i} \gamma^\mu Q_i) (\bar{N} \gamma_\mu N)$  & $~ $ 4-fermion & $\alpha^{(i)}_{QNN}$   \\ \cline{2-3}\cline{5-5} 
$~$ 4-fermion N & $\mathcal{O}^{(i)}_{LNN}$ ($\mathcal{O}_{\rm LN}^{\beta}$) & $(\bar{L_i} \gamma^\mu L_i) (\bar{N} \gamma_\mu N)$  & $~$  vector- mediated  & $\alpha^{(i)}_{LNN}$   \\ \cline{2-3}\cline{5-5} 
$ $ &  $\mathcal{O}^{(i)}_{fNN}$ ($\mathcal{O}_{\rm ff}$) &   $(\bar f_i \gamma^{\mu}f_i) (\bar N \gamma_{\mu}N)$  & $~$ $f=u, d, l$  & $\alpha^{(i)}_{fNN}$   \\ \hline
 $~ $ 4-fermion CC & $\mathcal{O}^{(i, j)}_{duNl}$ ($\mathcal{O}_{\rm duN\ell}^{\beta}$) &  $ (\bar d _j \gamma^{\mu} u _j) (\bar N \gamma_{\mu} l_i)$ & $ ~$ 4-fermion vector- mediated  & $\alpha^{(i, j)}_{duNl}= \alpha^{(i, j)}_{V_0}$    \\ \hline 
 $ $ & $\mathcal{O}^{(i, j)}_{QuNL} $ ($\mathcal{O}_{\rm QuNL}^\alpha$) & $(\bar Q _i u _i)(\bar N L_j)$ & $ ~$ 4-fermion & $ \alpha^{(i,j)}_{QuNL}=\alpha^{(i,j)}_{S_1}$  \\ \cline{2-3}\cline{5-5} 
$ ~$ 4-fermion &  $\mathcal{O}^{(i, j)}_{LNQd} $   ($\mathcal{O}_{\rm LNQd}^\alpha$)  & $(\bar L_i N) \epsilon (\bar Q _j d _j)$  & $ ~$ scalar-mediated  & $\alpha^{(i,j)}_{LNQd}=\alpha^{(i,j)}_{S_2}$ \\ \cline{2-3}\cline{5-5}
$~$  CC/NC & $\mathcal{O}^{(i, j)}_{QNLd}$ & $(\bar Q _i N)\epsilon (\bar L_j d_j)$  &  $ $ & $\alpha^{(i,j)}_{QNLd}= \alpha^{(i,j)}_{S_3}$ \\ \cline{2-3}\cline{5-5} 
$ $ & $\mathcal{O}^{(i, j)}_{LNLl}$  ($\mathcal{O}_{\rm LNL\ell}^{\delta\beta}$) & $(\bar L_i N)\epsilon (\bar L_j l_j)$  & $ $ & $\alpha^{(i, j)}_{LNLl}=\alpha^{(i, j)}_{S_0}$ \\ \lasthline
\end{tabular}
\end{adjustbox}
\caption{{Basis of $d=5$ and $d=6$ operators with a right-handed neutrino $N$ \cite{delAguila:2008ir, Liao:2016qyd}. Here $l_i$, $u _i$, $d _i$ and $L_i$, $Q _i$ denote the right-handed singlets and the left-handed $SU(2)$ doublets, respectively. The field $\phi$ is the scalar doublet, $B_{\mu\nu}$ and $W_{\mu\nu}^I$ are the $U(1)_{Y}$ and $SU(2)_{L}$ field strengths. Also $\sigma^{\mu \nu}=\frac{i}{2}[\gamma^{\mu}, \gamma^{\nu}]$ and $\epsilon=i\sigma^{2}$ is the anti-symmetric symbol in two dimensions. We follow the notation in \cite{delAguila:2008ir} and quote the names in \cite{Fernandez-Martinez:2023phj}.}\label{tab:Operators}}
\end{table}

In our simplified scenario, we assume that the mixings $U_{\ell N}$ from Eq.~\eqref{eq:mixing} are very small, so their effects on the heavy state $N$ can be neglected. Instead, we focus on the impact of new physics that could arise from heavy mediators at a higher energy scale $\Lambda$. These effects are captured by a set of effective operators  $\mathcal{O}_\mathcal{J}$ which are constructed using SM fields and the right-handed neutrino $N_R$, and are consistent with the SM gauge symmetry $SU(2)_L \otimes U(1)_Y$ gauge symmetry \cite{delAguila:2008ir, Liao:2016qyd, Wudka:1999ax}, and lead to the following Lagrangian:
\begin{eqnarray}\label{eq:Lagrangian}
\mathcal{L}=\mathcal{L}_{SM}+\sum_{d=5}^{\infty}\frac1{\Lambda^{d-4}}\sum_{\mathcal{J}} \alpha_{\mathcal{J}} \mathcal{O}_{\mathcal{J}}^{d}
\end{eqnarray}
where $d$ is the mass dimension of the operator $\mathcal{O}_{\mathcal{J}}^{d}$, $\alpha_{\mathcal{J}}$ are the effective (Wilson) couplings and the sum in $\mathcal{J}$ goes over all independent interactions at a given dimension $d$. The complete list of $\nu$SMEFT operators up to $d=6$ is given in Table \ref{tab:Operators}. Note that we do not consider the $d=5$ operators, as they do not contribute to the processes studied at lepton or electron-proton (eP) colliders -nor to the decay of the heavy $N$- when only one right-handed neutrino state is present and the heavy-active mixings $U_{\ell N}$ are neglected. We therefore focus solely on the contributions from $d=6$ operators, following the approach in \cite{delAguila:2008ir, Liao:2016qyd}. The implementation of the effective Lagrangian in \texttt{FeynRules 2.3} has been discussed in \cite{Zapata:2022qwo}. A detailed discussion of the role of each operator in the heavy $N$ phenomenology, the existing constraints on the Wilson coefficients and the full expressions for each explicit Lagrangian term can be found in \cite{Zapata:2023wsz}.  

Many of the operators in Table \ref{tab:Operators} can contribute directly to the heavy $N$ production in future lepton and eP colliders, as well as to its decay modes and total decay width value. While most of the recent works studying the $\nu$SMEFT phenomenology focus on the impact of specific operators \cite{Mitra:2024ebr, Barducci:2022hll, Beltran:2021hpq} or take them to act separately at a time \cite{Fernandez-Martinez:2023phj}, here we consider what we call an agnostic -or democratic- scenario, where we take into account the simultaneous effect of every dimension 6 operator in Table \ref{tab:Operators}. 

Our approach consists in taking the numerical value of the quotients $\alpha_{\mathcal{J}}/\Lambda^2$ to the same numerical value for every $\mathcal{J}$ in Eq.~\eqref{eq:Lagrangian}. This simplifies the parameter space to only two variables: the heavy neutral lepton mass $m_N$ and the effective interaction coefficients that weight the value of the physical observables $\frac{\alpha}{\Lambda^2}$. 

We follow this agnostic approach, since it leads to more realistic results, given that in most cases specific BSM UV models will generate not only one operator, but contribute to several interactions when the correct matching between scales for the model is calculated, due to operator mixing \cite{Chala:2020vqp, Beltran:2023ymm, Fernandez-Martinez:2023phj, Ardu:2024tzb}. 

This benchmark scenario implies that the total decay width of the heavy $N$ is calculated at tree level taking into account the interference terms between the contribution of each operator to each channel, as we did originally in our full calculation in \cite{Duarte:2015iba,Duarte:2016miz} and updated in \cite{Zapata:2023wsz}. This decay width -and every branching ratio- value is fixed for any given benchmark point in the $(m_N, \frac{\alpha}{\Lambda^2})$ plane. It is given as an input to the simulation software, what in turns allows us to impose constraints directly on a reduced two dimensional parameter space.

There are only two exceptions to this democratic numerical treatment:

\begin{itemize}

\item The most stringent limits that can be derived on the effective operators involve the first fermion family and come from the yet unobserved neutrinoless double beta decay ($0\nu\beta\beta$ decay). We consider every operator contributing to this process to be upper bounded by an $m_N$ dependent coefficient. We thus impose the value 
\begin{equation*}
\frac{\alpha_{0\nu\beta\beta}(m_N)}{\Lambda^2} = 3.2\times 10^{-8}\left(\frac{m_N}{100 ~ ~\rm{GeV}} \right)^{1/2}
\end{equation*}
on the coupling of every effective operator contributing to the vertex $u d N e$. The details of the derivation can be followed from \cite{Duarte:2016miz, Duarte:2014zea} and \cite{Beltran:2023ymm}. The couplings of the operators contributing to this vertex are: $\alpha^{(1)}_{N l \phi}\; $ and $\alpha^{(1)}_{NW} $ -which contribute through the interchange of a $W$ boson- and the four-fermion $\alpha^{(1, 1)}_{duNl}, ~\alpha^{(1, 1)}_{QuNL}, ~\alpha^{(1, 1)}_{LNQd} , ~\alpha^{(1, 1)}_{QNLd}$.

\item We consider a loop-factor $1/16\pi^2$ multiplying the couplings of the operators $\mathcal{O}_{NB}$ and $\mathcal{O}_{NW}$, which are generated at one loop-level in the UV theory \cite{delAguila:2008ir, Wudka:1999ax}, as shown in Table \ref{tab:Operators}.

\end{itemize}

On the other hand, the decision to take this agnostic point of view also implies that we must be careful when considering existing derivations for the bounds of the $\nu$SMEFT interactions. When more than one operator acting at a time is considered, bounds can be obtained from a variety of processes involving combinations of couplings, typically with different production and decay channels for the $N$. 

For instance, limits on the effective couplings can be obtained from the well known existing bounds on the seesaw mixings $U_{\ell N}$ \cite{Fernandez-Martinez:2023phj}. One can consider the relation 
\begin{equation}\label{eq:miXalf}
U_{\ell N}\simeq  \frac{ v^2}{2}\frac{\alpha^{(i=\ell)}_{N l \phi}}{\Lambda^2}
\end{equation}
to derive bounds for the bosonic charged current effective coupling from the mixings with flavor $\ell \equiv l_i$ in the case of the muon and tau families. We thus can translate the most stringent bound to date from the LHC experiments for the hundred- GeV $N$ mass range \cite{CMS:2022hvh, ATLAS:2023tkz, ATLAS:2024rzi}, which is the bound of  $|V_{\ell N}|^2\leq 0.07$ for $m_N=500$ GeV \cite{ATLAS:2024rzi}, and leads to the limit $\frac{\alpha^{(i=\mu)}_{N l \phi}}{\Lambda^2}\leq 8.74 \times 10^{-6} ~\text{GeV}^{-2}$. This value exceeds the range considered here by more than an order of magnitude and is therefore not included in our results plots.

Recent works on $\nu$SMEFT  interactions including $d=6$ operators have derived bounds for the different Wilson couplings values $\alpha_{\mathcal{J}}$ -or alternatively on the new physics scale $\Lambda$- given by existing experimental direct or indirect searches of BSM phenomena. Most of these constraints are applicable for $m_N$ masses below the range we consider in this work ($m_N \leq m_W$) and definitely do not apply for the agnostic benchmark scenario considered here \cite{Fernandez-Martinez:2023phj,Mitra:2022nri,Mitra:2024ebr,Barducci:2022hll,Ardu:2024tzb}.  We remark that for masses $m_N$ well below the electroweak symmetry breaking scale, it would be convenient to consider the low energy effective field theory $\nu$LEFT, with QED and QCD invariant operators with sterile neutrinos \cite{Chala:2020vqp, Datta:2020ocb, Ardu:2024tzb}. 

As discussed in the Introduction, adopting an agnostic approach precludes direct matching to a specific UV-complete model underlying the effective interactions\footnote{While the study of specific UV-completions is beyond the scope of the present work, we invite the interested reader to visit recent works on the phenomenology of concrete models leading to interactions parameterized by the $\nu$SMEFT: Ref. \cite{Cottin:2021lzz} for scalar leptoquark mediated physics, Refs. \cite{Deppisch:2019kvs, Chiang:2019ajm} for models with $Z'$ bosons and \cite{deVries:2024mla} for the Left-Right Symmetric Model.}. Nevertheless, it provides a simplified and general framework in which sensitivity projections can be interpreted in a broad and tractable manner, complementary to the constraints derived on single coefficients, which can only be reliably translated to new physics that induces a single operator.

\section{Future Collider sensitivities}\label{sec:ColliderSensitivities}

In this work we want to tackle the sensitivity reach to the agnostic $\nu$SMEFT scenario of future collider experiments, focusing on the regime where the heavy $N$ decays promptly into leptonic or semi-leptonic final states, induced by 4-fermion vector and scalar $d=6$ interactions \footnote{The $N$ decay length can be obtained by $d_{lab}= \beta \gamma \frac{\hbar c}{\Gamma_N}= \frac{s-m_N^2}{2 s~ m_N}\frac{1.975\times 10^{-16}}{\Gamma_N} \text{m}$, with $\sqrt{s}$, $m_N$ and $\Gamma_N$ in GeV. We have checked numerically that the largest value occurs for $m_N=75$ GeV and $\alpha/\Lambda^2= 0.1 \times 10^{-6} ~\text{GeV}^{-2}$, giving $d_{lab}= 7.78 \times 10^{-9} \text{m}$. Thus we find that for the whole parameter space explored in the present work, the $N$ decays promptly.}. 
Previous studies \cite{Zapata:2022qwo} pointed to the ability of lepton colliders to discover these interactions with the use of angular observables like forward-backward asymmetries, and now we present the first prospective exclusion plots based on the study of single $N$ production with pure leptonic or semi-leptonic decays at an $e^+ e^-$ collider. 

\begin{figure}[tbp]
\center{
{\includegraphics[width=\textwidth]{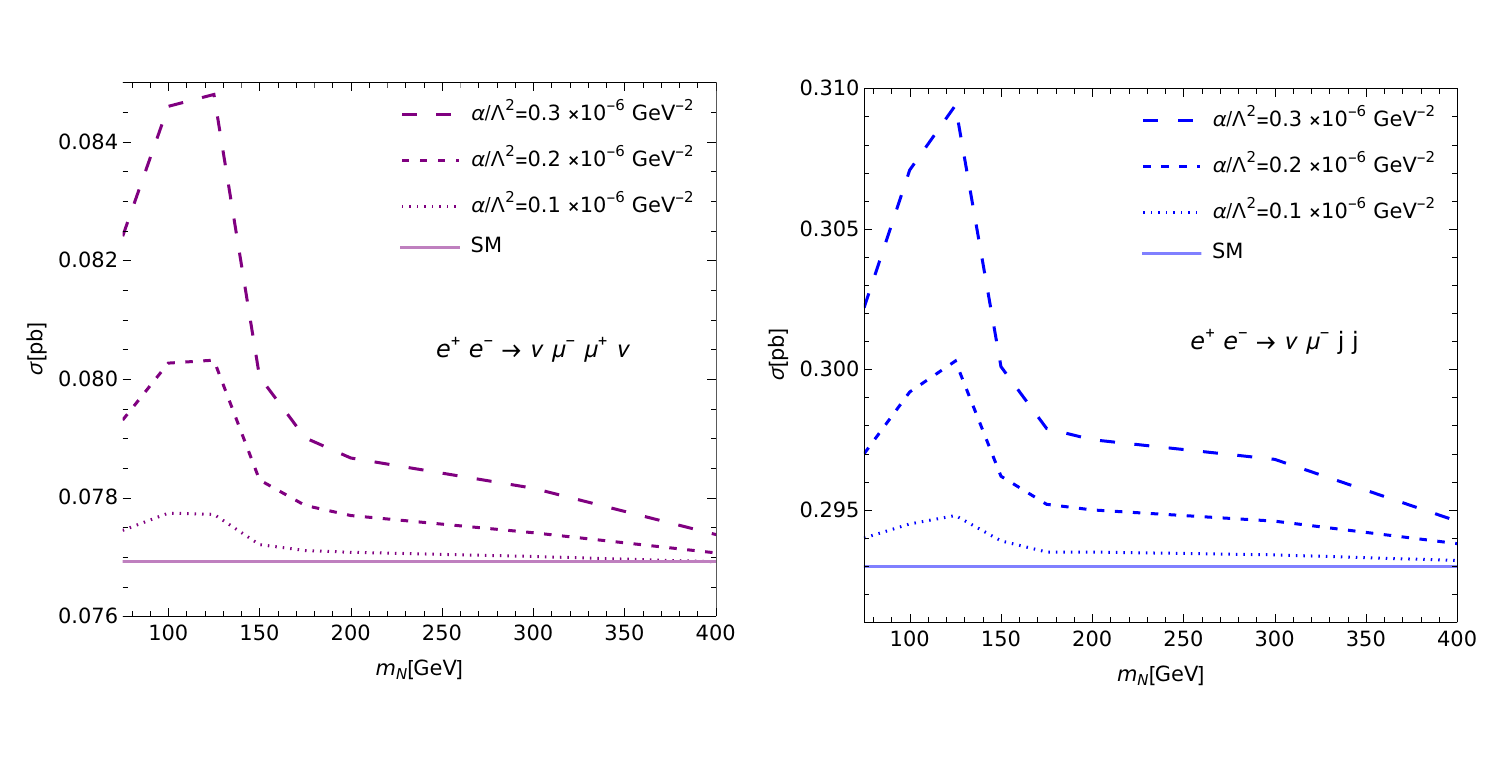}}
}
\caption{\label{fig:Xsect}{Cross section values for the benchmark points selected for the pure-leptonic (left) and semi-leptonic (right) channels, for $\sqrt{s}=500$ GeV.}}
\end{figure}

These signals can be studied in future lepton colliders like the linear ILC \cite{Bambade:2019fyw} or circular colliders like the FCC-ee \cite{FCC:2018byv} and the CEPC \cite{CEPCStudyGroup:2018ghi}. Here for simplicity we will consider an $e^{+}e^{-}$ collider with center of mass energy $\sqrt{s}=500 $~GeV and different integrated luminosities $\mathcal{L}$ for estimating the numbers of events.

The irreducible SM backgrounds for both the pure leptonic $e^{+}e^{-} \to \nu \mu^{-} \mu^{+} \nu $ and the semi-leptonic $e^{+}e^{-} \to \nu \mu^{-} \mathrm{j} \mathrm{j}$ processes, with $\nu=\nu_{e,\mu,\tau}$ and light flavored jets $\mathrm{j}= u,d,c,s$, involve diagrams with intermediate standard bosons -photons, $Z$- and Higgs bosons in $s$ channels, which subsequently decay into muon pairs, light neutrino pairs, or quark pairs, and $W$ bosons decaying leptonically or hadronically. The dominant SM backgrounds for both processes are events that come from $e^{+}e^{-} \to W^{-}W^{+}$, with both $W$ decaying leptonically in the first case \cite{Hernandez:2018cgc, Banerjee:2015gca}, and the $W^{+}$ decaying hadronically in the second \cite{Liao:2017jiz}.

We generate events in \texttt{MadGraph5\_aMC@NLO 3.4.1}  \cite{Alwall:2014hca, Alwall:2011uj}, producing LHE events at parton level, which are read by the embedded version of \texttt{PYTHIA 8}  \cite{Sjostrand:2014zea}, and then are interphased to \texttt{Delphes 3.5.0} \cite{deFavereau:2013fsa} with the DSiDi card \cite{Potter:2016pgp} for a fast detector simulation. The analysis of the generated events at the reconstructed level is made with the expert mode in \texttt{MadAnalysis5 1.8.58} \cite{Conte:2012fm}.

We adopt the following basic acceptance cuts for both the pure-leptonic and the semi-leptonic process: we keep transverse momenta for jets $p_{T}^{ \mathrm{j}} > 20$ GeV and leptons $p_{T}^{\ell} > 10 $ GeV, pseudorapidities $|\eta_{\mathrm{j}}|<5$, $|\eta_{\ell}|< 2.5$, and isolation between jets and leptons $\Delta R_{\mathrm{j} \mathrm{j}, \ell \ell, l \mathrm{j}} > 0.4$.

The cross section values for the selected benchmark points for both the pure leptonic and the semi-leptonic process are shown in Figure \ref{fig:Xsect}.

\begin{figure}[tbp]
\center{
{\includegraphics[width=0.5\textwidth]{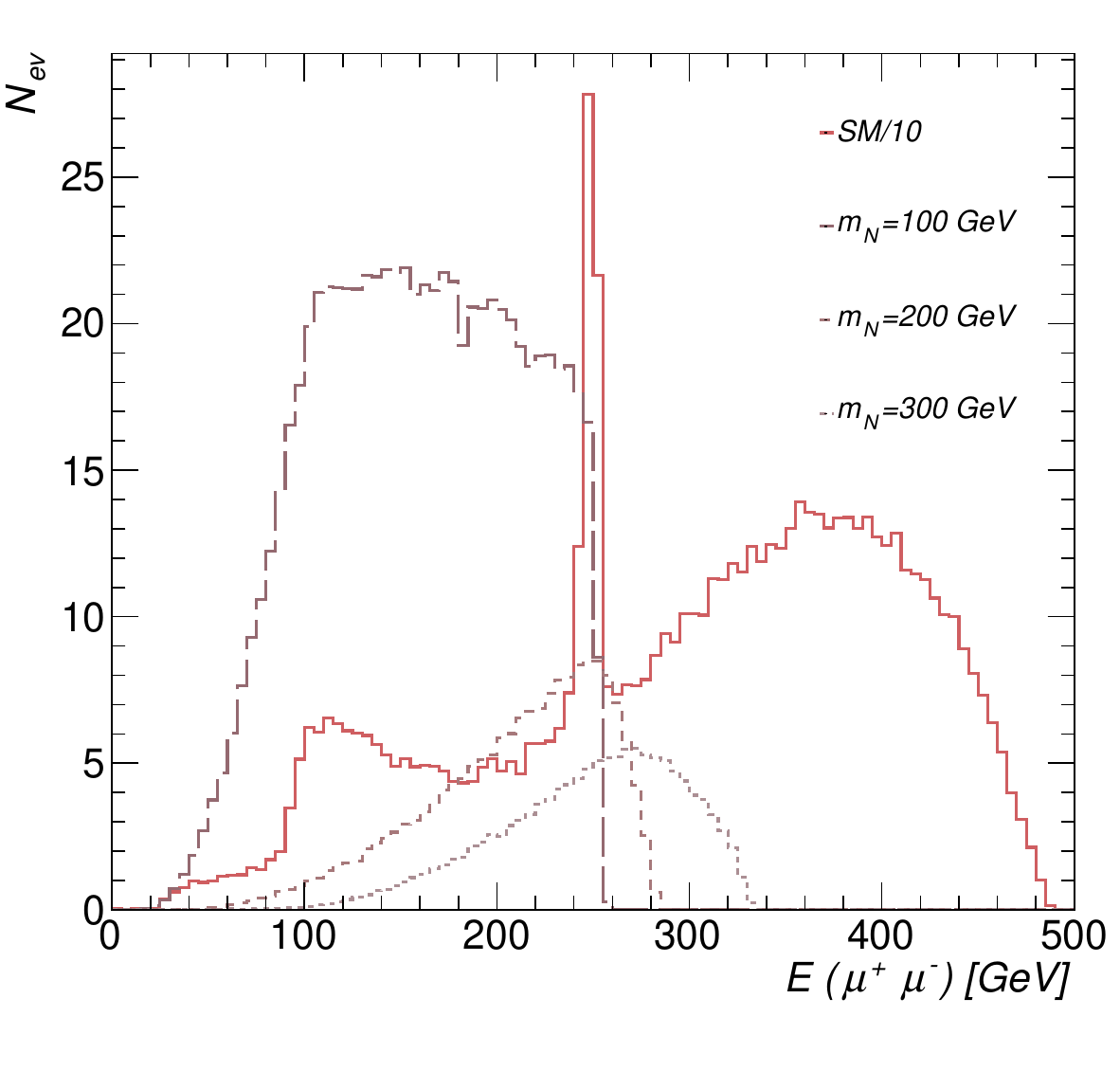}}~
{\includegraphics[width=0.5\textwidth]{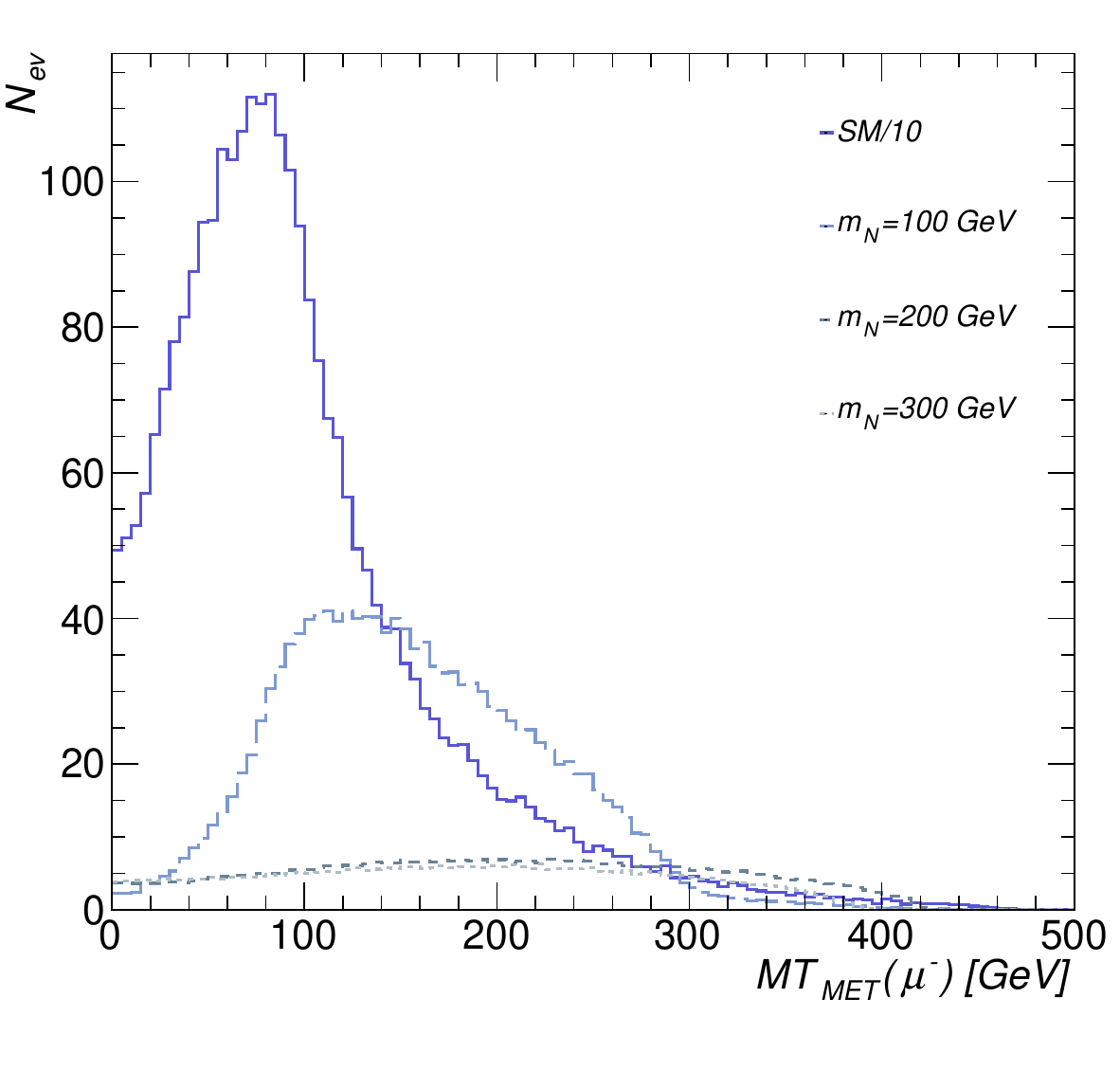}}
\caption{\label{fig:ee_cuts}{Distribution of the $E(\mu^{-}+\mu^{+})$ variable in $e^+e^- \to \nu N \to \nu \mu^- \mu^+ \nu$ (left) and the MT-MET($\mu$) variable in $e^{+}e^{-} \to \nu N \to \nu \mu^{-} \mathrm{j} \mathrm{j}$ (right) for pure signal processes with $\alpha/\Lambda^2= 0.3 \times 10^{-6} ~\text{GeV}^{-2}$ and the SM backgrounds, for $\mathcal{L}=100 ~\text{fb}^{-1}$ and $\sqrt{s}=500$ GeV.}}
}
\end{figure}

\paragraph{Pure-leptonic channel.}

We consider the process $e^+e^- \to \nu N \to \nu \mu^- \mu^+ \nu$, with $\nu=\nu_{e,\mu,\tau}$, which gives as a signal two opposite-sign muons and missing energy. The production vertex receives contributions from both the $\mathcal{O}_{N l\phi}$($\mathcal{O}_{\rm HN\ell}$) charged current operator and the scalar-mediated four lepton operator $\mathcal{O}_{LNLl}$. The pure-leptonic $N$ decay depends on explicit contributions from the same operators\footnote{We allow for the production of light neutrinos of every flavor $\nu_{\ell}$, with $\ell= e, \mu, \tau$, both in the $N$ production vertex and decay channels. This means we consider both Lagrangian terms generated by the $\mathcal{O}^{(i,j)}_{LNLl}$ operator: $\bar \nu_{L,i}N_R \bar \ell_{L,j} \ell_{R,j}$ and $\bar \ell_{L,i}N_R \bar \nu_{L,j} \ell_{R,j}$, fixing $j=2$ to consider muons, and summing over the allowed possibilities for $i$ given by each term. Also, we include $\mathcal{O}^{(i)}_{N l \phi ~(HNl)}$, which produces a $W$ boson. In the production vertex only the $i=1$ flavor contributes, while for the decay of the $N$ to $\mu^-$ and $\bar{\nu}_{\mu}$, the only contribution from this operator corresponds to $i=2$.}.

As the $N$ is produced together with a light neutrino in a 2-2 process, its energy and boost are completely determined in the c.m. frame for each mass value $m_N$. Its production is reflected in the dependence of the various observables on the summed energy of the di-muon pair $E_{\mu \mu}=E(\mu^{-}+\mu^{+})$. We produce events in a realistic dataset, generated with the SMNeff6 UFO \cite{Zapata:2022qwo} allowing for effective-SM interference (S+B) and separate them from the SM-only (B) events for the pure leptonic process $e^{+}e^{-} \to \nu \mu^{-} \mu^{+} \nu $. Informed by our previous work, we ask for a muon and an anti-muon in the final state, and apply a selection cut keeping events with a minimum value of missing transverse energy (MET > $25$ GeV) and keep events with $E_{\mu \mu}< 240$ GeV to separate the signal from the dominant SM background, which peaks at $E_{\mu \mu}= \sqrt{s}/2$, in a symmetric configuration where the muon anti-muon pair shares half the energy with the unobservable light neutrinos. In Figure \ref{fig:ee_cuts} (left) we show the distribution of this variable on events generated for the SM background and for pure-signal events -where an $N$ is indeed produced- for three different $m_N$ values and $\alpha/\Lambda^2= 0.3 \times 10^{-6} ~\text{GeV}^{-2}$, an integrated luminosity $\mathcal{L}=100 ~\text{fb}^{-1}$ and $\sqrt{s}=500$ GeV.

\paragraph{Semi-leptonic channel.} 

In the semi-leptonic process $e^{+}e^{-} \to \nu N \to \nu \mu^{-} \mathrm{j} \mathrm{j}$, again with $\nu=\nu_{e, \mu, \tau}$ the $N$ decays into a muon and two  light-flavored jets. While its production mechanism remains unchanged, the decay of the $N$ now involves the contributions of the vector four-fermion operator $\mathcal{O}^{(2,i)}_{duNl}$ together with the vector $\mathcal{O}^{(2)}_{Nl\phi}$ and the scalars $\mathcal{O}^{(2, i)}_{QuNL}$, $\mathcal{O}^{(2, i)}_{LNQd}$ and $\mathcal{O}^{(i, 2)}_{QNLd}$, which give two quarks (jets $\mathrm{j}$) in the final state \footnote{For a recent work on the same signal see \cite{Mekala:2023kzo}.}. 

As before, we allow for interference of the signal and the SM contributions, and perform a cut-based analysis, asking for a muon in the final state and selecting events with missing transverse energy (MET) greater than $25$ GeV to account for the final state neutrino. The transverse mass of the final muon-neutrino system, calculated in the reconstructed level data as the transverse mass (MT) of the missing transverse energy (MET)- muon system: MT-MET($\mu$) \footnote{The transverse mass variable of the muon-missing transverse energy system is defined as MT-MET($\mu$)= $\sqrt{2 p_T^{\mu} p_T^{miss}[1-cos(\Delta \phi(\vec{ p_T^{\mu}}, \vec{p_T^{miss}})]}$, we use the name given in \texttt{MadAnalysis5}.} allows to separate the signal events from the SM background, which peaks at the $m_W$ value, reflecting the fact that the muon-neutrino pair comes from a $W^{-}$. We find that a cut selecting events with MT-MET($\mu$)> 85 GeV keeps more than $70 \%$ of the events for the different $m_N$ signal datasets, while keeping near $10\%$ of the SM events. In Figure \ref{fig:ee_cuts} (right) we show the distribution of the MT-MET($\mu$) variable for the SM background and pure signal events, again for three different $m_N$ values and $\alpha/\Lambda^2= 0.3 \times 10^{-6}~ \text{GeV}^{-2}$, an integrated luminosity $\mathcal{L}=100 ~\text{fb}^{-1}$ and $\sqrt{s}=500$ GeV.

In order to obtain the exclusion plots for the agnostic $\nu$SMEFT parameter space, we generate events for different values of the couplings $\frac{\alpha}{\Lambda^2}$ and masses $m_N$ in a grid, also considering different total integrated luminosities $\mathcal{L}$ to estimate the number of signal and background events. 

The 95\% CL exclusion limits in the $(m_N, \frac{\alpha}{\Lambda^2})$ plane are calculated following the PDG review on Statistics \cite{ParticleDataGroup:2020ssz} and Appendix B in \cite{Magill:2018jla}. For each signal point, we calculate the upper number of signal events $s^{up}$ consistent at 95\% CL with the observation of the expected number of background events, by supposing that the data collected in the experiment exactly matches the integer part of the number of events for the background prediction. The shaded areas correspond to the parameter space regions where the interpolated expected number of signal events exceeds the upper allowed value $s^{up}$, and thus the limits are imposed directly on the agnostic $\nu$SMEFT parameter space values.

\begin{figure}[tbp]
\center{
{\includegraphics[width=0.5\textwidth]{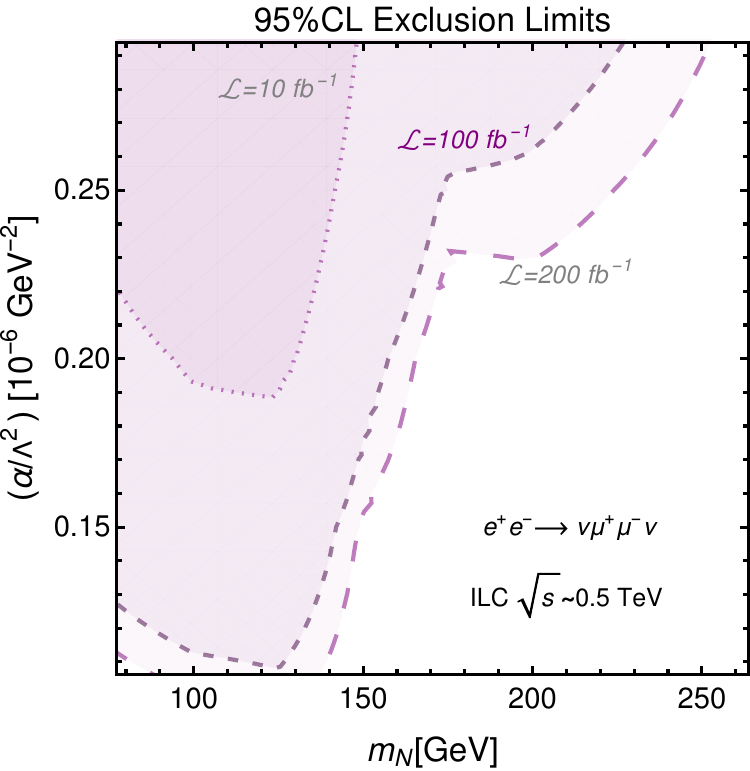}}~
{\includegraphics[width=0.5\textwidth]{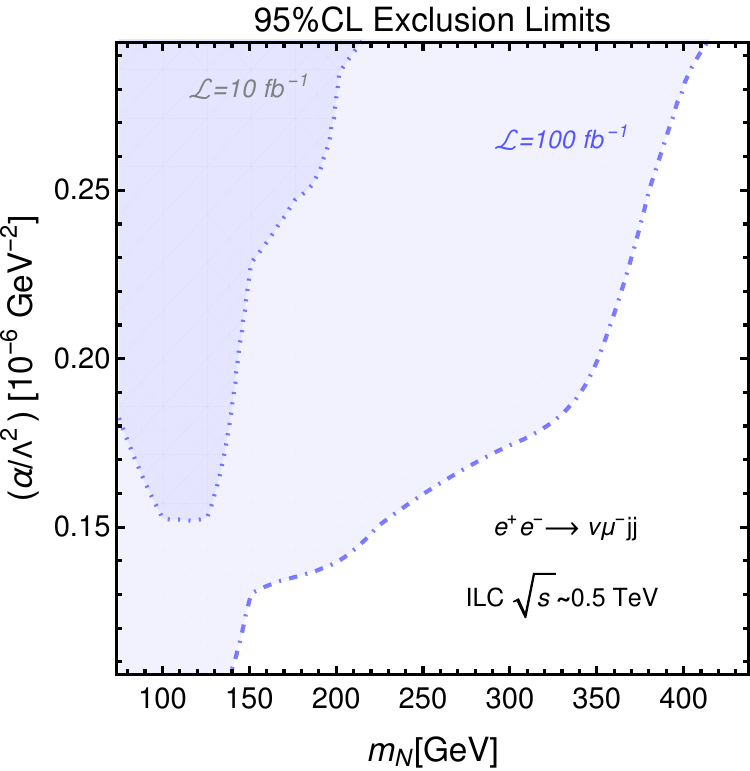}}
}
\caption{\label{fig:ee-muX}{95\% CL Exclusion limits reachable at $\sqrt{s}=0.5$ TeV for the pure-leptonic (left) and semi-leptonic (right) channels, for different values of integrated luminosity $\mathcal{L}$.}}
 \end{figure}

Figure \ref{fig:ee-muX} shows the projected 95\% CL exclusion limits in the $(m_N, \frac{\alpha}{\Lambda^2})$ plane for the pure-leptonic (left) and semi-leptonic (right) processes at a future electron-positron collider for the agnostic $\nu$SMEFT benchmark scenario. Our results indicate that, in both decay channels, these experiments could start constraining the effective couplings for masses $m_N$ below $200$ GeV, even with an integrated luminosity of $\mathcal{L} = 10~\rm{fb}^{-1}$. With a luminosity of $100~\rm{fb}^{-1}$, the sensitivity reach of a lepton collider exploiting the semi-leptonic channel can test $N$ masses as large as $500$ GeV.

This sensitivity reach can be compared to the sensitivity found for lepton-trijet processes at an electron-proton collider like the future LHeC, already obtained for the $\nu$SMEFT agnostic benchmark scenario in \cite{Zapata:2023wsz}. Those limits were obtained for the LHeC \cite{LHeC:2020van, AbelleiraFernandez:2012cc, Bruening:2013bga}, with a center-of mass energy close to $1.3$ TeV, and an integrated luminosity $\mathcal{L}=100 ~\rm fb^{-1}$, performing a BDT analysis with the \texttt{Root TMVA} package \cite{Hocker:2007ht}. 

\begin{figure}[tpb]
\centering
{\includegraphics[width=0.6\textwidth]{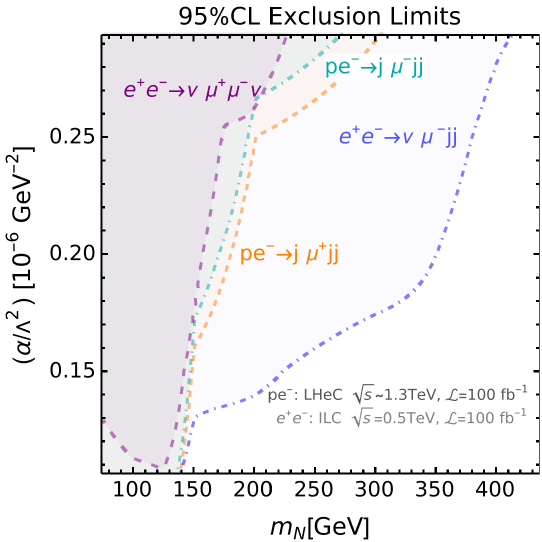}}
 \caption{ \label{fig:ExcluAll}{{95\% CL exclusion limits for the agnostic $\nu$SMEFT at future colliders.}}}
 \end{figure}

In Figure \ref{fig:ExcluAll} we plot the expected exclusion limits for the lepton number conserving -but lepton flavor violating- muon-trijet signal $pe^{-} \to \mathrm{j} \mu^- \mathrm{j} \mathrm{j}$ in cyan, and the lepton number and flavor violating $pe^{-} \to \mathrm{j} \mu^+ \mathrm{j} \mathrm{j}$ in orange, together with the lepton collider expected sensitivities already shown in Figure \ref{fig:ee-muX} in purple and blue, all for an estimated number of events calculated with the same integrated luminosity $\mathcal{L}=100 ~\rm fb^{-1}$. We find that the four channels studied could exclude the lower mass region ($m_N \lesssim 130~\rm{GeV}$) with couplings $\frac{\alpha}{\Lambda^2}> 10^{-7}~\rm{GeV}^{-2}$. The sensitivity reach falls for higher $m_N$ values, due to the decrease of the branching ratio of $N$ to fermions when it can start decaying to on-shell Higgs bosons \cite{Zapata:2022qwo,Zapata:2023wsz}. 

\section{Summary and Conclusions}\label{sec:summary}

In this work we show the first prospective 95\% CL exclusion plots reachable at a future lepton collider with a c.m. energy of $\sqrt{s}=0.5 ~\rm{TeV}$ for what we call the agnostic $\nu$SMEFT scenario, in the high mass regime where the heavy $N$ can decay promptly to leptons and jets that can easily be measured in detectors. We obtain the excluded regions in the $\frac{\alpha}{\Lambda^2}$ vs. $m_N$ plane. Figure \ref{fig:ee-muX} shows that the semi-leptonic channel $e^+e^-\to \nu N \to\nu \mu^- \mathrm{j} \mathrm{j}$ can reach the highest sensitivity, due in principle to the bigger expected  cross section \cite{Zapata:2022qwo} and an efficient cut-based analysis proposed to separate signal from SM backgrounds.     

We also compare the obtained limits with those found for the reach of the LHeC eP collider in Figure \ref{fig:ExcluAll}. Both machines would be able to study the electroweak $N$ mass regime with interaction couplings in the ballpark of $\frac{\alpha}{\Lambda^2} \sim 10^{-7} \rm{GeV^{-2}}$. These effective couplings limits are as low as the ones that are considered in recent works for the lower mass regime where the $N$ can be long-lived and be found in displaced decay searches, both at the LHC and future colliders. One can find sensitivity prospects for the near-future experiments concerning the dimension-6 $\nu$SMEFT interactions in \cite{Beltran:2021hpq} for a long-lived $N$ at the LHC exploiting possible displaced vertices searches, and in \cite{Barducci:2022hll} for prompt and displaced $N$ decays at future Higgs factories, in both cases for lighter $N$ benchmark scenarios with $m_N \lesssim 60 ~\text{GeV}$. Also, a variety of testable signals in planned experiments are discussed in \cite{Mitra:2022nri}, and constraints on the $pp \to e N$ cross section interpreted from recent LHC searches are obtained in \cite{Mitra:2024ebr} for vectorial $\nu$SMEFT interactions. Our group is currently working on a detailed recast of LHC searches for heavy neutral leptons in terms of the agnostic $\nu$SMEFT benchmark scenario. 

The discovery of heavy neutrinos would have profound implications on the current landscape of high energy physics and our understanding of Nature; however, constraining the potential new physics underlying neutrino mass generation could also provide a pathway to uncovering the origin of the observed neutrino masses -one of the most significant unresolved questions in particle physics-.

{\bf{Note added:}} When the initial manuscript was under revision, a new preprint appeared \cite{Bolton:2025tqw}, which studies sensitivity prospects for some $\nu$SMEFT operators for the FCC-ee lepton collider with $\sqrt{s}= 91.2 ~\text{and} ~240$ GeV, for mono-photon and missing energy as well as displaced vertices searches, considering operators to act separately at a time. In Figure 12, it shows a comparison with previous bounds and prospects -from LEP, PMNS and lower energy experiments- displaying their results in black lines. The benchmark parameter space we presented here only covers the portion of $m_N> 75$ GeV in those plots. However, it can clearly be seen that for the studied operators, their prospective bounds are in the ballpark of those obtained here for the agnostic scenario for higher center of mass energies, with values of $c_i/\Lambda^2 \sim 10^{-7} \text{GeV}^{-2}$. As the authors mention, they show results for Dirac neutrinos, which avoid the most restricting constraints from neutrinoless double beta decay for the first flavor family. We thus consider our results to be nicely complementary, suggesting that a more detailed study of the signals proposed here could also lead to competitive sensitivities. 

\acknowledgments

We thank PEDECIBA (Uruguay) for the financial support.

\appendix

\bibliographystyle{bibstyle}
\bibliography{Bib_N_4_25}

\end{document}